\documentclass{iaus}
\usepackage{graphicx}
\def\eg{{\it e.g.,\ }}

\title[Planetary Nebulae and Stellar Populations] %% give here short title %%
{Planetary Nebulae as Probes of Stellar Populations}

\author[Ciardullo]   %% give here short author list %%
{Robin Ciardullo}

\affiliation{Department of Astronomy \& Astrophysics, Penn State University,
University Park, PA 16802, USA \break email:rbc@astro.psu.edu}

\pubyear{2006}
\volume{234}  %% insert here IAU Symposium No.
\pagerange{1--8}
\date{1 May 2006}
\jname{Planetary Nebulae in Our Galaxy and Beyond}
\editors{M.J. Barlow \& R.H. M\'endez}
\begin{document}

\maketitle

\begin{abstract}
Planetary nebulae (PNe) have the potential to revolutionize our 
understanding of extragalactic stellar populations.  Indeed, in many systems, 
bright PNe are the only individual objects identifiable from the ground, and,
even more often, they are the only stars that are amenable to spectroscopy.
We review the ways in which ensembles of PNe can be used to probe the 
metallicity, age, and history of a stellar population.  In particular,
we discuss three cases:  the weak line spectroscopic regime, where one has 
knowledge of the line-strengths of faint forbidden lines such as 
[O~III] $\lambda 4363$, a bright-line regime, where only the strongest
emission lines are visible, and the photometric regime, where the only
information available is the [O~III] $\lambda 5007$ luminosity function.
We show that each of these cases, when properly calibrated, can provide
unique insights into the objects that make up a stellar population.

\end{abstract}

\firstsection % if your document starts with a section,
              % remove some space above using this command.
\section{Introduction}
In order to understand how galaxies are put together, we must be able to
characterize the collective properties of their stars.  Information
about a system's age (or history of star formation), metallicity (or
history of chemical enrichment), initial mass function, and binarity are all
critical for modeling the processes which define galactic evolution.
Unfortunately, due to the great distances involved, obtaining such data
is often extremely difficult.

This is where planetary nebula (PN) research can help.  All stellar 
populations older than $\sim 10^8$~yr make planetary nebulae, and, in a large
galaxy, hundreds of PNe are available for study.  More
importantly, PNe are bright.  In systems outside the Local Group, PNe 
are often the only individual stars that can be identified from the ground, 
and are almost always the only (non exploding) stars that can be studied 
spectroscopically.   Thus, planetary nebulae can provide 
unique insights into the history of a stellar system.

The amount of information provided by a population of PNe depends on the type 
of data that one can acquire.  In nearby systems, it is possible to use 4-m and
8-m class telescopes to observe all the important emission lines produced by
a PN, including the critical diagnostic lines of [O~III] $\lambda 4363$,
[S~II] $\lambda\lambda 6716,6731$, and [N~II] $\lambda 6548, 6584$.  With 
this, one can directly measure a population's abundance distribution function 
and constrain its history of chemical enrichment. In more distant galaxies, 
these weak emission features are usually unavailable, but one can still use 
brighter lines such as [O~III] $\lambda 5007$, [Ne~III] $\lambda 3869$, 
H$\alpha$, and H$\beta$ to probe stellar content.  Finally, if PN spectroscopy 
is impossible, one can still investigate galactic history via the [O~III] 
$\lambda 5007$ planetary nebula luminosity function (PNLF).  We discuss 
each of these regimes below.

\section{The Weak Line Regime}
A decade ago, measurements of the critically important [O~III] $\lambda 4363$
emission line were mostly restricted to the PNe of the Galaxy and the 
Magellanic Clouds.  For example, it was only through heroic efforts that 
\cite{richer} and \cite{jc99} were able to detect this feature in a sizeable
sample (42) of PNe in M31's bulge, and thereby measure the population's oxygen
abundance.  These authors obtained the interesting result that the metallicity 
spread for M31's bulge stars is over a dex, and that the region's mean oxygen 
abundance is slightly less than solar.  Unfortunately, given the limitations 
of the 4-m apertures that were available at the time, follow-up studies proved 
to be impossible.

Today in this era of 8-m class telescopes, it is possible to obtain
abundances for hundreds of PNe throughout the largest of the Local Group
galaxies, M31 and M33.  With these data, one can characterize the systems'
abundance distribution functions, not only for oxygen, but for the 
$\alpha$-process elements of neon, sulfur and argon, all of which are 
unaffected by stellar evolution.  Moreover, because one knows the distances to
these extragalactic PNe, the combination of [O~III] $\lambda 5007$
photometry and spectrophotometry can be used to derive the objects'
{\it absolute\/} line fluxes.  This is important, because from this
information, it is possible to deduce the locations of the central stars 
on the HR diagram, estimate their masses (via comparisons with 
post-AGB evolutionary tracks; \cite[Vassiliadis \& Wood 1994]{vw94}),
and infer the masses of the stars' progenitors (via the initial mass-final 
mass relation; \cite[Weidemann 2000]{weidemann}).  In other words, deep 
spectrophotometry of a large number of planetary nebulae should allow us 
to trace the chemical enrichment of a stellar population back through time.

Although such an analysis is exciting, it is also extremely difficult.
For example, consider the PNe of the Large Magellanic Cloud.  Excellent
spectrophotometry and nebular models exist for $\sim 60$~objects from
the work of Dopita \etal\ (1991a,b; 1997).  If one
uses these data to derive $\alpha$-element abundances, and central star
luminosities, temperatures, and (ultimately) masses, then a clear trend 
emerges:  PNe from lower-mass (older) progenitors are systematically more 
metal-poor than their younger counterparts.  This suggests that we can, 
indeed, use PNe to probe a galaxy's history of chemical enrichment.  However, 
if one instead uses ultraviolet measurements of the nebular and stellar 
continuum to place the central stars on the HR diagram (\cite[Herald \& 
Bianchi 2004]{herald}; \cite[Villaver \etal\ 2003]{villaver03}; 
\cite[Stasi\'nska \etal\ 2004]{stas}), then much of this correlation goes away. 
Clearly, in order to exploit PNe to their fullest, our inferences about 
central star properties need to be made more robust.

\section{The Bright-Line Regime}
Once outside the Local Group, the [O~III] $\lambda 4363$ emission line
of planetary nebulae becomes extremely hard, if not impossible, to detect.
Although this makes traditional nebular modeling impossible, it does not
preclude the use of PNe for population studies.  Indeed, \cite{djv92} and
\cite{n4697} have shown that the ratio of [O~III] $\lambda 5007$
to H$\beta$ can be used to place useful constraints on a PN's oxygen abundance.

But one can do better than that.  Due to the poor distances of Galactic PNe,
samples of Milky Way planetaries are extremely heterogeneous in nature.
With high and low-core mass PNe mixed together, it is nearly impossible to
separate out planetaries at their peak luminosity from PNe that have already
faded by several magnitudes.  In extragalactic systems, this is not a 
problem.  Since the absolute luminosities of all the PNe are known,
it is possible to define statistically complete samples of objects at the
bright end of the luminosity function and be assured that all are in a 
similar phase of evolution.  For example, PNe in the Milky Way 
exhibit a wide range of excitation, with [O~III] $\lambda 5007$ to 
H$\alpha$ emission line ratios ($R$) ranging from $\sim 4$ to $\sim 0.1$. 
However, when one studies extragalactic PNe, a pattern emerges:  the 
brightest PNe {\it all\/} have $R > 2$, regardless of their parent
stellar population (\cite[Ciardullo \etal\ 2002]{p12}).

In fact, the best way to analyze the bright emission lines from a 
homogeneous PN dataset is through a differential, 
multi-dimensional analysis.  A schematic example of this 
approach is shown in Figure~1.  In the figure, PNe of the LMC 
(from \cite[Meatheringham \& Dopita 1991a,b]{md1,md2}) and 
NGC~4697 (\cite[M\'endez \etal\ 2005]{n4697}) are plotted in a
3-dimensional space that has the absolute [O~III] $\lambda 5007$ line flux on 
the $y$-axis, and the [O~III]/H$\beta$, and [Ne~III] $\lambda 3869$/H$\beta$ 
line ratios on the $x$- and $z$-axes.  The differences in the populations are
obvious.  With a reasonable amount of effort, this type of data could 
be acquired for a range of stellar populations out to $\sim 10$~Mpc, and serve
as the basis for an investigation of PN production and evolution as a function
of galactic environment.

\begin{figure}
\centerline{
\scalebox{0.55}{
\includegraphics{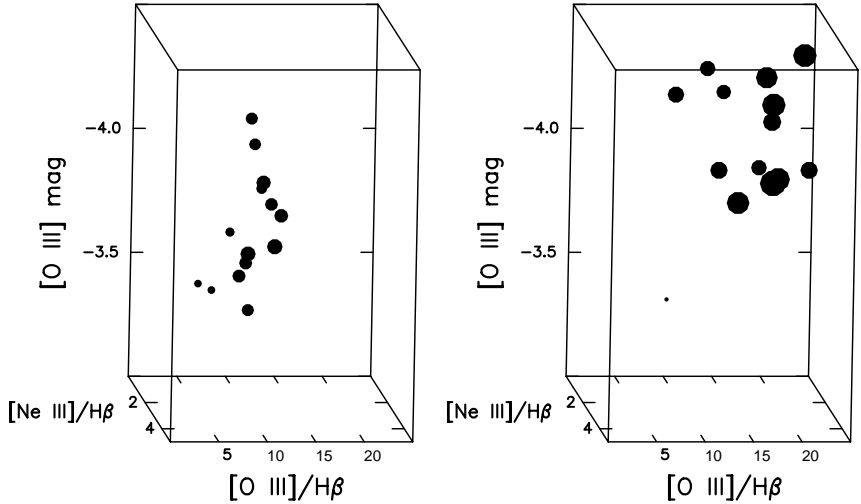}
}}
\caption[]{The PNe of the LMC (left) and the large elliptical NGC~4697 (right)
plotted in a three-dimensional phase space, with the absolute [O~III]
magnitudes displayed on the $y$-axis.  Note how this type of plot clearly
distinguishes the different PN populations.}
\end{figure}

\section{The Photometric Regime}
Spectroscopy of extragalactic PNe is time consuming, and in many cases
impossible.  Hence one often must rely on [O~III] $\lambda 5007$
(or possibly H$\alpha$) photometry to probe a PN population.
Observations in over 50 galaxies have shown that the bright-end of the
[O~III] $\lambda 5007$ planetary nebula luminosity function
is well-described by the truncated exponential
\begin{equation}
N(M) \propto e^{0.307 M} \left\{ 1 - e^{3(M-M^*)} \right\} 
\end{equation}
where
\begin{equation}
M = -2.5 \log F_{5007} - 13.74
\end{equation}
and $M^* = -4.47$ is the absolute magnitude of the most luminous planetary
(\cite[Jacoby \etal\ 1992]{mudville}; \cite[Ciardullo \etal\ 2002]{p12}).
Since $M^*$ is an excellent standard candle
for all but the most metal-poor systems, its brightness cannot be used for
stellar population studies.  However, two properties of the PNLF
do hold promise for population measurements.

The first of these is a ``dip'' in the luminosity function which starts
$\sim 2$~mag below $M^*$.  This feature is not present in every galaxy:  
in M31, the faint-end of the PNLF is well described by the simple 
\cite{hw63} exponential, which is expected from an ensemble of uniformly 
expanding spheres ionized by non-evolving central stars 
(\cite[Ciardullo \etal\ 1989, 2002]{p2,p12}).  However, as
Figure~2 illustrates, star-forming systems such as the Small Magellanic Cloud 
and M33, have a factor of $\sim 2$ fewer PNe than predicted from the 
exponential law (\cite[Jacoby \& De Marco 2002]{jdm02}; \cite[Ciardullo 
\etal\ 2004]{m33}).  Clearly, this part of the PNLF is sensitive to some 
aspect of stellar population.

\begin{figure}
\centerline{
\scalebox{0.5}{
\includegraphics{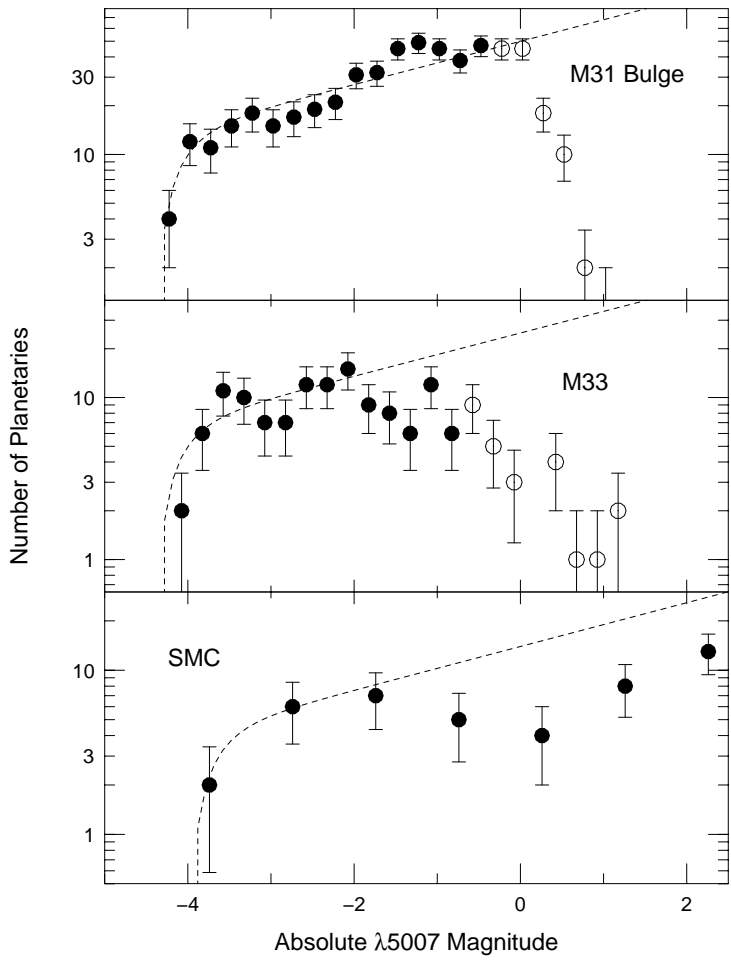}
}}
\caption{The [O~III] $\lambda 5007$ planetary nebula luminosity functions
for M31's bulge, M33's disk, and the Small Magellanic Cloud.  Note the 
differences between the populations:  star forming systems have a strong 
deficit of intermediate-brightness planetaries.}
\end{figure}

The second population dependent feature of the [O~III] PN luminosity
function is its normalization.  Following \cite{p2}, this property is usually
parameterized by $\alpha$, the number of PNe detected above a certain
absolute magnitude (typically 0.5 or 2.5~mag below $M^*$), divided by the
total (bolometric) luminosity of the stellar parent population.  In the 
star-forming galaxies of the Local Group, $\alpha$ is in good agreement
with the predictions of stellar evolution models (\cite[Buzzoni,
Arnaboldi, \& Corradi 2006]{buzzoni}), but in older stellar populations,
this is not always the case.  In fact, some elliptical galaxies have 
almost an order of magnitude fewer PNe than one might expect.  
This variation has the potential to be an extremely powerful 
probe of early-type systems.

\section{Explaining the PNLF}
To date, there has been no attempt to simultaneously model both the shape and
normalization of the PNLF\null.  Moreover, those models which have concentrated
solely on the shape of the luminosity function have failed spectacularly, 
either by requiring a constant star-formation rate for elliptical galaxies 
(\cite[M\'endez \& Soffner 1997]{mendez}) or predicting a bright-end cutoff 
that is several magnitudes fainter than what is observed 
(\cite[Marigo \etal\ 2004]{marigo}).  Obviously, an important piece of physics 
is missing from these analyses.

The main difficulty in modeling the PNLF comes from the absolute magnitude
of its cutoff.  An $M^*$ PNe emits $\sim 600 L_{\odot}$ in its [O~III]
$\lambda 5007$ emission line.  Since no more than $\sim 10\%$ of the
central star's flux can come out in this line, the luminosity of the source
powering the nebula must be at least $\sim 6,000 L_{\odot}$.  Unfortunately,
the only post-AGB cores capable of generating this much energy are those with
masses of $M > 0.6 M_{\odot}$ (\cite[Vassiliadis \& Wood 1994]{vw94}), and 
these require progenitor stars with $M \gtrsim 2 M_{\odot}$ 
(\cite[Weidemann 2000]{weidemann}).  Such stars have lifetimes of less than 
a Gyr, and are not present in any number in elliptical galaxies 
(\cite[\eg Trager \etal\ 2005]{trager}).

As Figure~3 demonstrates, the situation is even worse than it looks.  
Spectroscopic observations in the Large Magellanic Cloud 
(\cite[Meatheringham \& Dopita 1991a,b]{md1,md2})
and M31 (\cite[Jacoby \& Ciardullo 1999]{jc99}) demonstrate that
internal extinction is extremely important in [O~III]-bright planetary 
nebulae.  While no PN is {\it observed\/} to have an [O~III] luminosity 
brighter than $M^*$, many PNe have intrinsic [O~III] fluxes that are much more 
luminous than this, sometimes by a factor of $\sim 2$.  Such a high 
luminosity can only be generated by a massive post-AGB core.

\begin{figure}
\centerline{
\scalebox{0.6}{
\includegraphics{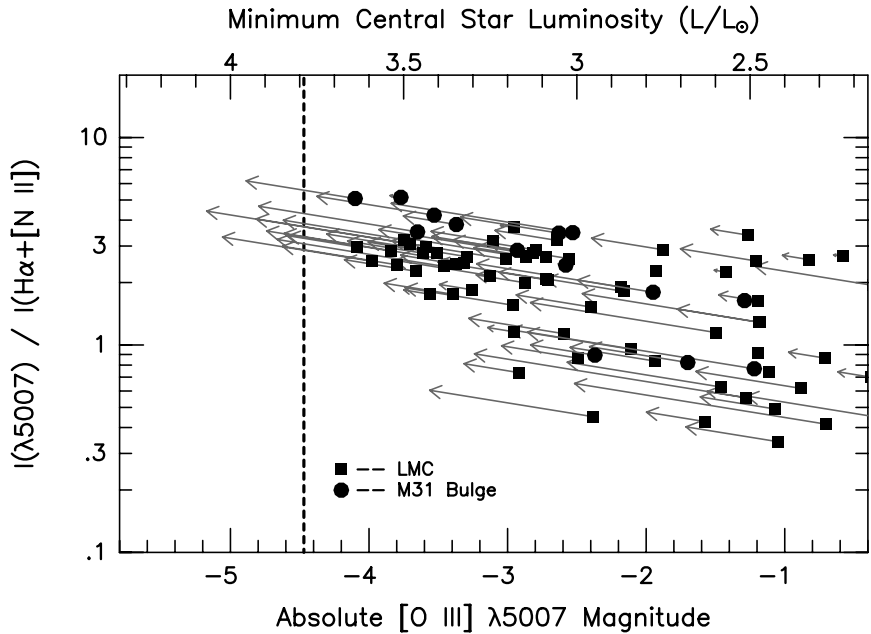}
}}
\caption{The observed ratio of [O~III] $\lambda 5007$ to H$\alpha$ 
plotted against [O~III] $\lambda 5007$ absolute magnitude for those bright PNe 
in the LMC and M31 with published spectrophotometry.  The arrows display 
the effects of internal extinction, as estimated from the objects' Balmer 
decrement.  The location of the PNLF cutoff, $M^*$, is identified via the 
dotted line.  Note that, although no PN is ever {\it observed\/} to have 
an [O~III] $\lambda 5007$ magnitude brighter than $M^*$, many have intrinsic 
luminosities much brighter than this.}
\end{figure}

In star-forming systems, the presence of high core-mass central stars
is not a problem.  But what produces these objects in old stellar populations?
The number of PNe observed in a galaxy, normalized to the galaxy's underlying
(bolometric) luminosity is given by
\begin{equation}
\alpha = B \, \cdot \, \tau \, \cdot \, f 
\end{equation}
where $B$ is the population's luminosity specific stellar evolution
flux, $\tau$, the lifetime of the PN stage, and $f$, the fraction of 
stars which evolve through the PN phase.  The value of $B$ is known
from the theory of stellar energy generation:  all old stellar populations, 
regardless of their age, metallicity, or initial mass function, have
$B \sim 2 \times 10^{-11}$~stars~yr$^{-1}~L_{\odot}^{-1}$
(\cite[Renzini \& Buzzoni 1986]{renzini}).  The lifetime, $\tau$, is 
similarly well-constrained, as models show that [O~III]-bright
PNe spend $\sim 500$~yr in the top 0.5~mag of the luminosity function 
(\cite[Marigo \etal\ 2004]{marigo}).  Therefore, observed values of 
$\alpha$ in the range $\sim 5 \times 10^{-10} < \alpha_{0.5} <
3 \times 10^{-9}$~PNe~$L_{\odot}^{-1}$ (\cite[Ciardullo \etal\ 2005]{c05})
translate into stellar fractions of between 5\% and 30\% (see Figure~4).
Since absorption line spectroscopy of early-type galaxies place the 
fraction of young stars far below this value 
(\cite[Trager \etal\ 2005]{trager}), the bright PNe 
of these systems cannot be evolving via normal (single star) stellar
evolution.  We must look elsewhere for their explanation.

\begin{figure}
\centerline{
\scalebox{0.7}{
\includegraphics{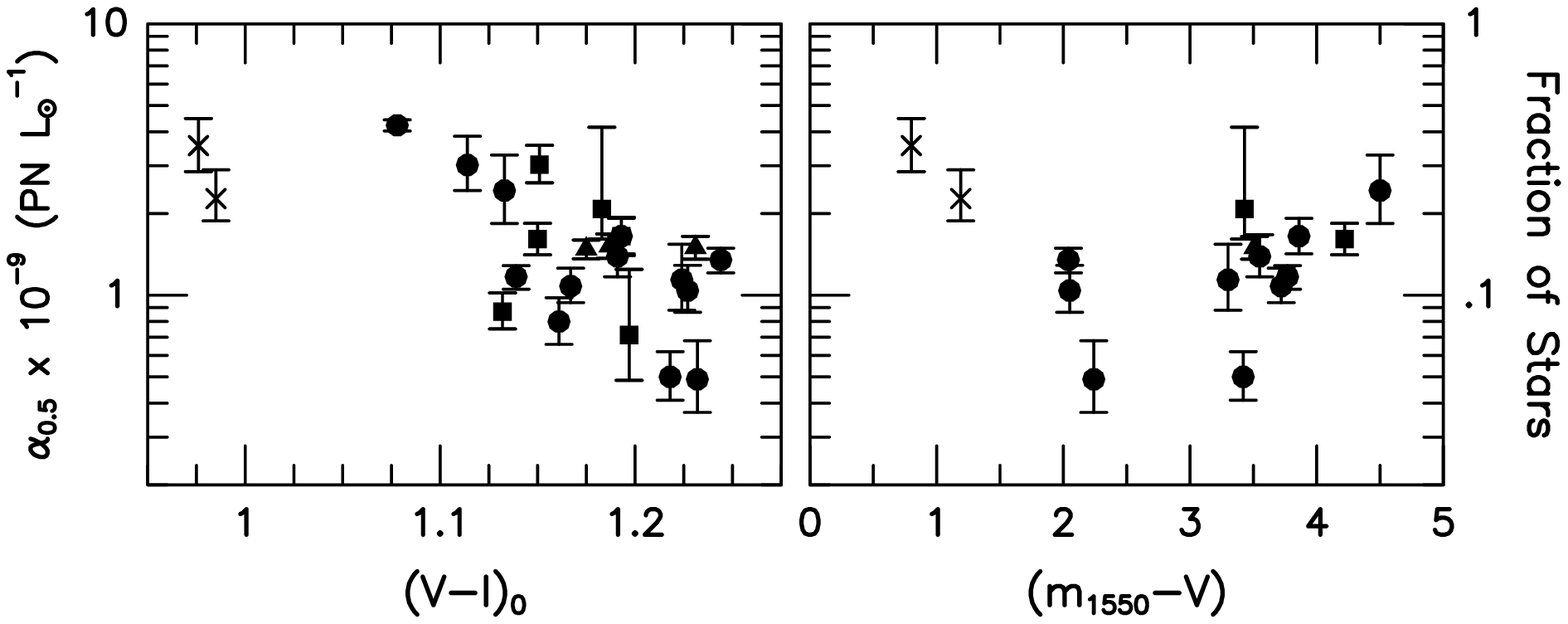}
}}
\caption{The bolometric-luminosity specific number of PNe in the top 
0.5~mag of the [O~III] PNLF plotted against the parent galaxy's 
optical and ultraviolet color.  Circles indicate elliptical galaxies, squares
are lenticulars, triangles are spiral bulges, and crosses represent
systems which contain obvious evidence of star formation.  The error
bars represent formal $1 \, \sigma$ uncertainties and are lower limits to the
true errors.  The right hand axis translates the PN densities to the
fraction of stars turning off the main sequence which evolve into
[O~III]-bright planetaries, under the assumption that the PNe are
products of single star evolution.}
\end{figure}

As of this date, two scenarios have been proposed to explain the
existence of [O~III]-bright PNe in early-type galaxies.  The first, described 
by \cite{c05}, involves 
binary coalescence and blue stragglers.  Although elliptical galaxies 
possess few stars with $M \sim 2 M_{\odot}$, they do contain many stars with
$M \sim 1 M_{\odot}$.  If some of these stars come into contact while still in
their hydrogen burning phase, the result may be conservative 
\cite{mccrea}-style mass transfer and coalescence.  This mechanism, which
has been associated with the creation of blue stragglers 
(\cite[Mateo \etal\ 1990]{mateo}; \cite[Carney \etal\ 2001]{carney}),
has the potential to build the requisite high mass cores, even in systems
as old as $\sim 10$~Gyr.

The blue straggler hypothesis has several advantages.  The first, and
most obvious, is that such stars are known to exist.  In fact, if blue
stragglers live $\sim 5 \times 10^8$~yr 
(\cite[Lombardi \etal\ 2002]{lombardi}), then the ratio of [O~III]-bright 
PNe to blue stragglers in systems such as the Ursa Minor dwarf
galaxy and the Milky Way halo is approximately the same as that of the
objects' lifetimes (\cite[Ciardullo \etal\ 2005]{c05}).  This is consistent
with the idea that one object evolves into the other.  Moreover, the binary 
merger hypothesis provides a natural explanation for the observed inverse 
correlation between $\alpha$ and the excess ultraviolet flux often associated
with elliptical galaxies.  As a stellar system ages, its main sequence turnoff 
mass decreases, and this decreases the mass of its post-AGB cores 
(\cite[Weidemann]{weidemann}).  In the extreme, these low-mass cores will
evolve too slowly to form planetary nebulae, and instead produce naked
UV-bright sources (\cite[Greggio \& Renzini 2000]{greggio}).  Meanwhile, the
smaller turnoff mass also reduces the number systems capable of merging
into stars above the critical $\sim 2 M_{\odot}$ threshold.  
Consequently the number of $M^*$ planetaries will also drop.  If this 
scenario is correct, then $\alpha$ can be used in conjunction with a 
galaxy's UV upturn to probe the precise location of a population's 
main-sequence turnoff, even when this turnoff is far below the limit 
of detectability.

The blue straggler hypothesis only runs into problems when one considers
the constraints it imposes on the integrated spectra of elliptical galaxies.
As pointed out by \cite{xin}, a sizeable population of blue stragglers can
mimic the presence of young stars, and lead to a significant underestimate
for the age of the stellar population.  In fact, \cite{buzzoni}
suggest that any blue straggler contribution to an old stellar system
must be considerably less than $\sim 10\%$, in order to avoid measurable 
effects on the galaxy's colors.   However, whether this is a serious problem
for models of elliptical galaxies is still an open question.

An alternative scenario for the presence of $M^*$ [O~III] $\lambda 5007$ 
sources in early-type populations involves symbiotic stars.  \cite{soker}
has proposed that mass transfer from a proto-planetary post-AGB star
onto a companion white dwarf can produce the energy and emission lines
needed to mimic a bright planetary nebula.  This model, which has
many commonalities with models of super-soft x-ray sources (albeit with a
much higher accretion rate), has the advantage that it does not
stress population synthesis predictions:  since the lifetime of this accretion
phase can be much longer than $\sim 500$~yr, far fewer [O~III] systems are
needed to populate the bright end of the luminosity function.  However,
the model has no natural explanation for the similarity between the maximum
brightness of a true planetary nebula and one of these symbiotic sources,
nor does it explain the inverse correlation between $\alpha$ and 
a population's UV upturn.

If there is indeed a second population of objects that evolve through
a non-traditional (binary-star) mode of stellar evolution, then the 
dip in the PNLF has a simple explanation.  High core mass planetaries,
whether from single star or binary evolution, will always form a luminosity 
function that has a distinctive dip at intermediate magnitudes.  The only
way to avoid such a feature is to fill in the dip with lower mass objects.
The amplitude of the dip, along with its precise location, would then be a
measure of how many of these difficult-to-observe low-mass stars are
present in the population and contributing to the PN luminosity
function.  If this idea is correct, then the agreement between M31's PNLF and 
the \cite{hw63} exponential law is purely coincidental.

\begin{acknowledgments}
This research made use of the NASA Extragalactic Database and was supported
in part by NSF grants AST 00-71238 and AST 03-02030.

\end{acknowledgments}

\end{document}